\documentclass[
superscriptaddress,
 amsmath,amssymb,
 aps,
 prl,
 twocolumn,
floatfix,
longbibliography
]{revtex4-2}

\usepackage{graphicx}
\usepackage{dcolumn}
\usepackage{setspace}
\usepackage{bm}
\usepackage{multirow}
\usepackage[colorlinks,linkcolor=blue,anchorcolor=blue,citecolor=blue,urlcolor=blue]{hyperref}
\usepackage[utf8]{inputenc}
\usepackage{graphicx}
\usepackage{threeparttable}
\usepackage{multirow,booktabs}
\usepackage{orcidlink}

\newcommand{\Ecm}{E_{\rm c.m.}}

\begin{document}
\title{
In search of beyond mean-field signatures in heavy-ion fusion reactions}

\author{R.~T. deSouza\,\orcidlink{0000-0001-5835-677X}}\email[Corresponding author: ] {desouza@indiana.edu}
\affiliation{Department of Chemistry and Center for Exploration of Energy and Matter, Indiana University, Bloomington, Indiana 47408, USA}

\author{K. Godbey\,\orcidlink{0000-0003-0622-3646}} 
\email{godbey@frib.msu.edu}
\affiliation{
FRIB Laboratory, Michigan State University, East Lansing, Michigan 48824, USA} 

\author{S. Hudan\,\orcidlink{0000-0002-9722-2245}}
\email{shudan@indiana.edu}
\affiliation{Department of Chemistry and Center for Exploration of Energy and Matter, Indiana University, Bloomington, Indiana 47408, USA}

\author{W. Nazarewicz\,\orcidlink{0000-0002-8084-7425}}\email{witek@frib.msu.edu}
\affiliation{Department of Physics and Astronomy and FRIB Laboratory, Michigan State University, East Lansing, Michigan 48824, USA}

\date{\today}

\begin{abstract}
Examination of high-resolution, experimental fusion excitation functions for $^{16,17,18}$O + $^{12}$C  reveals a remarkable irregular behavior 
that is rooted in the structure of both the
colliding nuclei and the quasi-molecular composite system. 
The impact of the
$\ell$-dependent fusion barriers is assessed using a time-dependent Hartree-Fock 
model. Barrier penetrabilities, taken directly from a  
density-constrained calculation, provide a significantly improved
description of the experimental data as compared to the standard Hill-Wheeler
approach. The remaining deviations between the parameter-free theoretical mean-field
predictions and experimental fusion cross sections are exposed and discussed.
\end{abstract}

\maketitle
The merging of two nuclei can provide a window into nuclear dynamics on short timescales. Heavy-ion fusion is governed by the interaction of the colliding nuclei resulting from the  delicate time-dependent balance of the repulsive electrostatic force and the  attractive nuclear force in the presence of angular momentum
for non-central collisions. Of fundamental importance in describing heavy-ion fusion is the collective potential of the two colliding nuclei, collective excitations of projectile and target, and the appearance of clustering effects during the fusion process. Progress in experiment, theory, and high performance computing   allows a direct confrontation  of  high-resolution fusion measurements  with advanced time-dependent theoretical frameworks to provide new insights into fusion dynamics. 

{\it Experimental evidence.---}  Indirect evidence for the transient configurations in fusion was first provided by examination of elastic scattering in $^{12}$C + $^{12}$C \cite{Bromley60}. Irregular energy dependence of the elastic cross-section  was interpreted as the formation of ``molecular states'' at specific energies. This  behavior was attributed to the deformability of the carbon nuclei \cite{Vogt60}. Absence of such behavior in $^{16}$O + $^{16}$O \cite{Bromley60} was interpreted in terms of the reduced deformability of the tightly bound, doubly-magic $^{16}$O nucleus \cite{Vogt60}.
A direct examination of the fusion excitation function  for $^{12}$C + $^{12}$C \cite{Sperr76}, $^{16}$O + $^{12}$C \cite{Sperr76a, Kovar79}, $^{16}$O + $^{16}$O \cite{Fernandez78, Kolata79}, and 
$^{20}$Ne + $^{20}$Ne \cite{Tserruya78} reveals the presence of an oscillatory structure in the near-barrier regime. This zigzag structure can be understood as originating from the accumulation of cross-section associated with successive individual $\ell$-waves with slightly different barriers \cite{Esbensen12, Wong12, Simenel13,Rowley15}. In order to directly probe the existence of transient configurations, particularly those that are weakly populated, it is crucial to disentangle the underlying macroscopic contribution due for example to 
$\ell$-wave dependent barriers. 
In the present work, we 
utilize high-resolution experimental data to confront state-of-the-art time-dependent Hartree-Fock (TDHF) calculations. 

High-resolution fusion excitation functions were obtained both by using recent active-target measurements as well as by combining prior thin-target measurements. Fusion was identified either by the direct detection of the heavy fusion products following de-excitation or by their secondary $\gamma$-emission. Any contribution from breakup prior to fusion, expected to be small for the energies and systems considered in this work, is not accounted for. Obtaining these high-resolution excitation functions was the key first step in this work.

\begin{figure}[htb]
\includegraphics[width=1.0\columnwidth]{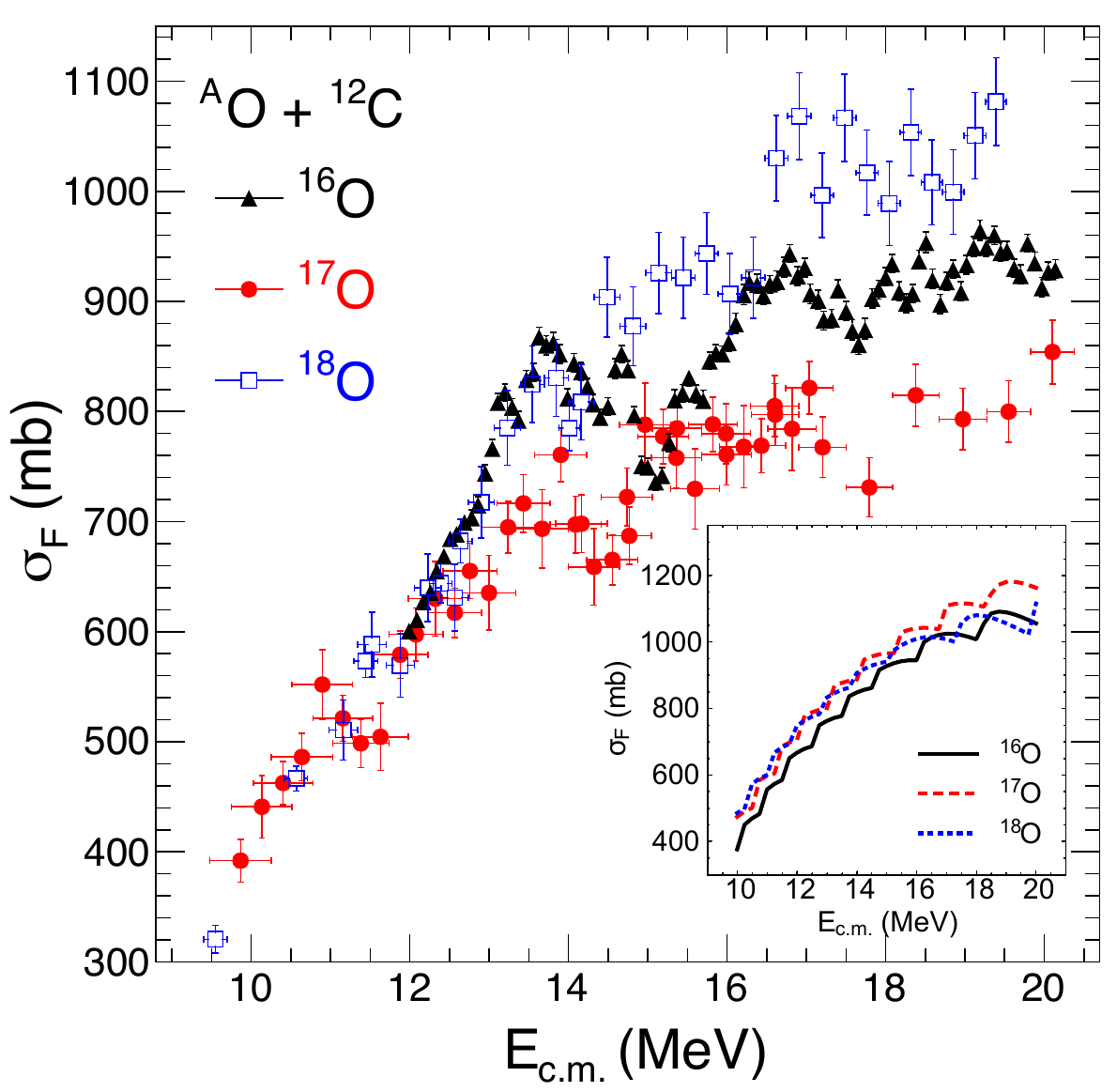}
\caption{ 
Experimental fusion excitation functions for the reactions of $^{16}$O (black triangles) \cite{Frawley82}, $^{17}$O (red dots) \cite{Hudan23}, and $^{18}$O (open squares) \cite{Johnstone22,Steinbach14a} impinged on a $^{12}$C target. 
The inset shows the results of TDHF$^*$. See text for details.
}
\label{fig:Fig1}
\end{figure}

Comparison of fusion processes induced by $^{16,17,18}$O nuclei
provides insight into three highly interesting cases. The $^{16}$O represents the reference case of a doubly-magic, tightly-bound nucleus. In the case of $^{17}$O, an odd unpaired neutron   occupies the $0d_{5/2}$ shell, resulting in a ground-state spin 5/2$^+$.  The extent to which this valence neutron is strongly or weakly coupled to the core is expected to impact  the fusion cross-section.  In the case of $^{18}$O, the two valence neutrons  form a Cooper pair. Pairing correlations are expected to impact the fusion cross section in two ways: by increasing the fusion barrier and by enhancing the neutron pair transfer. The experimental excitation functions for $^{16,17,18}$O + $^{12}$C are presented in Fig.~\ref{fig:Fig1}.

Direct comparison of these three experimental excitation functions alone provides considerable information. While the excitation functions exhibit common features, notable differences exist. All the excitation functions shown 
in Fig.~\ref{fig:Fig1} manifest a zigzag behavior superimposed on the overall increase in cross-section with increasing energy. Significantly more structure is observed for $^{16}$O with prominent peaks observed at $\Ecm \approx 11$\,MeV, 14\,MeV, and 16.5\,MeV. The magnitude of these peaks is reduced for $^{17}$O and  $^{18}$O. At  lower energies, all the excitation functions are rather similar suggesting that in this regime the valence neutrons in $^{17,18}$O play a spectator role. In contrast, the reduction in cross-section for $^{17}$O as compared to $^{16}$O at higher energies is particularly noteworthy. If the valence neutron in $^{17}$O is weakly coupled to the $^{16}$O core one might expect either an increased fusion cross-section  due to an increased spatial extent of the neutrons or essentially no increase at all if neutron breakup preceded fusion. The reduction of the fusion cross-section for $^{17}$O thus suggests that in this energy regime the presence of the valence neutron does influence fusion. This influence could be associated with the increased role of breakup and neutron transfer which can suppress the above-barrier cross-section while enhancing the below-barrier cross-section \cite{Gollan21}. The  enhanced fusion cross section at $\Ecm>14$\,MeV for $^{18}$O as compared to $^{16}$O suggests that pairing correlations impact the fusion cross section  at higher energies.

\begin{figure}
\includegraphics[width=1.0\columnwidth]{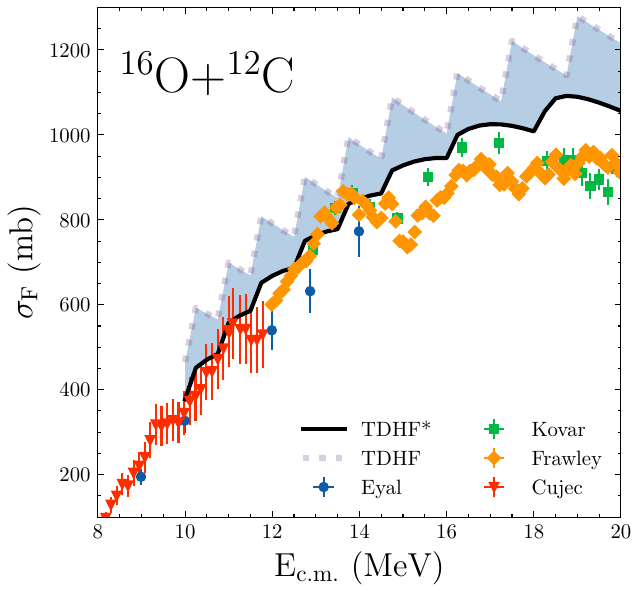}
\caption{ 
Comparison of experiment with theory for the fusion excitation function for the $^{16}$O+ $^{12}$C reaction . Experimental data are taken from Refs.   \cite{Eyal76} (blue circles), \cite{Kovar79} (green squares), \cite{Frawley82} (orange diamonds), and \cite{ Cujec76} (red upside-down triangles). Raw TDHF results are shown with a light dotted line and modified DC-TDHF/TDHF hybrid results are shown with a solid black line. The difference between TDHF and TDHF$^*$ is highlighted by shading.
}
\label{fig:Fig2}
\end{figure}

In order to provide the most complete, high-resolution description of the fusion excitation function for $^{16}$O + $^{12}$C several datasets have been combined and the result is
presented in Fig.~\ref{fig:Fig2}. The cross-section at higher energies which relies on the direct detection of the fusion products \cite{Kovar79, Eyal76, Frawley82} is augmented by indirect measurements of the cross-section at lower incident energies \cite{Cujec76}. Measurement of fusion at higher incident energies that relied on $\gamma$-ray measurements were excluded due to larger uncertainties. 
The  reported cross-sections depicted in  Fig.~\ref{fig:Fig2} are internally very consistent. The high resolution data not only reveals the peaks in the cross-section at $\Ecm \approx 11$\, MeV, 14\,MeV, and 16.5\,MeV previously noted but also an oscillatory behavior at lower energies. 

{\it Theoretical framework.---} To understand the fusion excitation functions, we have performed TDHF calculations for the above-barrier collisions.
On general grounds, a TDHF approach is well suited to describe the large-amplitude collective motion associated with fusion while also describing the transfer dynamics, equilibration processes, and Pauli blocking that affect heavy-ion fusion probabilities~\cite{godbey2017, simenel2020, simenel2017}. 

Recently, advances in theoretical and computational techniques have allowed TDHF calculations to be performed on a three-dimensional (3D) Cartesian grid thus eliminating artificial symmetry restrictions~\cite{Keser12}. The unrestricted 3D geometry allows for precise simulations that can capture the rich time-dependent dynamics at play in light nuclear reactions~\cite{simenel2013a,godbey2019b}.
Although in the sub-barrier regime it is necessary to perform density constrained TDHF (DC-TDHF) calculations \cite{Umar12, deSouza13} to obtain the heavy-ion potentials \cite{Umar2006,Simenel13}, at the above-barrier energies considered in this work direct TDHF calculations can be performed by initiating collisions for a series of increasing impact parameters until the maximum impact parameter for fusion is reached. Moreover, the barrier associated with each incoming  $\ell$-wave can be determined by finding the lowest energy associated with each $\ell$-window. 
This collision energy was scanned in steps of 0.25\,MeV across the reported range of energies for all systems. The effective interaction represented by energy density functional (EDF) used in this work was primarily UNEDF1~\cite{kortelainen2012}, though a set of parameters chosen from the Bayesian posterior distribution~\cite{mcdonnell2015} was also used to assess the sensitivity of the reaction outcomes to the choice of EDF~\cite{godbey2022}.
The same systematic calculations were performed  for all three oxygen beams. For the $^{18}$O reaction the frozen pairing approximation was employed, as in Ref.~\cite{Steinbach14a}. In contrast to the variations seen in fusion studies for heavier nuclei~\cite{reinhard2016,godbey2022}, the above-barrier fusion cross sections have been found to be largely insensitive to the choice of effective interaction.
While the unrestricted 3D Cartesian geometry affords a more flexible computational framework, it comes at an increased cost with each simulation requiring a few hours on a standard multicore compute node.
For the entire study, considering three systems, around 3000 individual trajectories were simulated to precisely determine the capture cross sections across a wide range of impact parameters and energies above the barrier.
Illustrative videos of the time evolution of the neutron  localization function \cite{Schuetrumpf2017} obtained in our TDHF simulations can be found in 
the Supplemental Material\,\cite{sup}. 

The fusion cross section can be expressed as
\begin{equation}
\sigma =\frac{\pi\hbar^2}{2\mu E_{\rm{c.m.}}}{\sum_{\ell=0}^{\ell_{\rm max}}
 (2\ell+1)P_{\ell}},
\end{equation}

where $\mu$ is the reduced mass, $E_{\rm{c.m.}}$
is the center-of-mass energy, 
$P_\ell$ is the probability of the $\ell$-wave fusing, and $\ell_{\rm max}$ corresponds to the largest $\ell$-wave that fuses.
For the raw TDHF results, $P_\ell$ is $1$ if the system fuses and $0$ if it does not.

The TDHF calculations were performed for $8<\ell\leq 20$. For
each $\ell$, a sharp increase in cross section is observed when the barrier for that particular $ \ell$-wave is surpassed. Tunnelling through the barrier mitigates this sharp threshold behavior \cite{Simenel13,Rowley15}. While the Hill-Wheeler approximation is often used for the penetrability, this approach presumes the transmission through an inverted parabolic potential. This assumption becomes progressively worse with increasing $\ell$-wave, particularly as $\ell$ approaches $\ell_{\rm max}$. In the current work, we extract $P_\ell$ directly from the penetrability of the computed DC-TDHF potentials for that $\ell$ value thus providing a self-consistent microscopic approach. In the event that $\ell_{\rm max}$ is different between the TDHF and DC-TDHF approaches, the lower of the two is chosen. In the following, we refer to this method as the  hybrid DC-TDHF/TDHF approach and designate it TDHF$^*$. The primary difference between TDHF$^*$ and the standard treatment for TDHF as detailed in Refs.~\cite{Simenel13,Rowley15} is that the cross sections are suppressed in addition to having a smoother behavior.

{\it Discussion.---}The  predictions of the  TDHF* model for the three reactions considered is shown in the inset of Fig.~\ref{fig:Fig1}. As might be naively expected from geometrical considerations based on mass scaling, $^{16}$O exhibits a smaller cross-section than $^{17,18}$O. The predicted trend differs from that of the experimental data shown in  Fig.~\ref{fig:Fig1}.

\begin{figure}[htb]
\includegraphics[width=1.0\columnwidth]{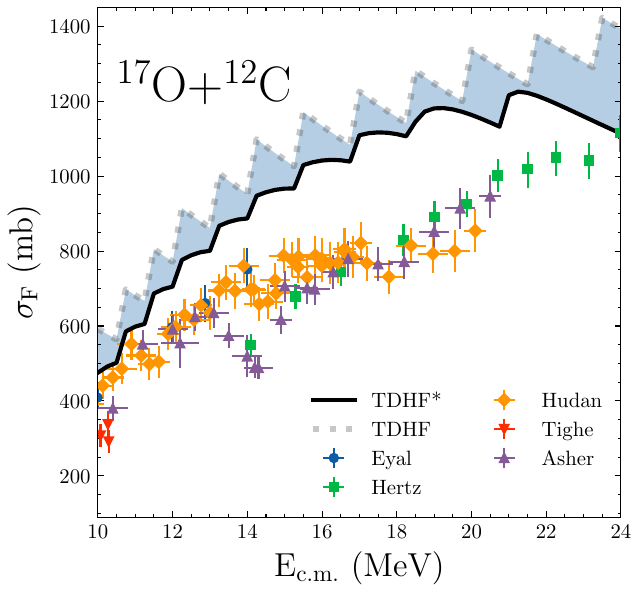}
\caption{Similar as in Fig.\,\ref{fig:Fig2} but for
the $^{17}$O+ $^{12}$C reaction. Experimental data are taken from Refs. \cite{Eyal76} (blue circles), \cite{Hertz78} (green squares), \cite{Hudan23} (orange diamonds), \cite{Tighe93} (red upside-down triangles), and \cite{Asher21a} (purple triangles). 
}
\label{fig:Fig3}
\end{figure}

\begin{figure}
\includegraphics[width=1.0\columnwidth]{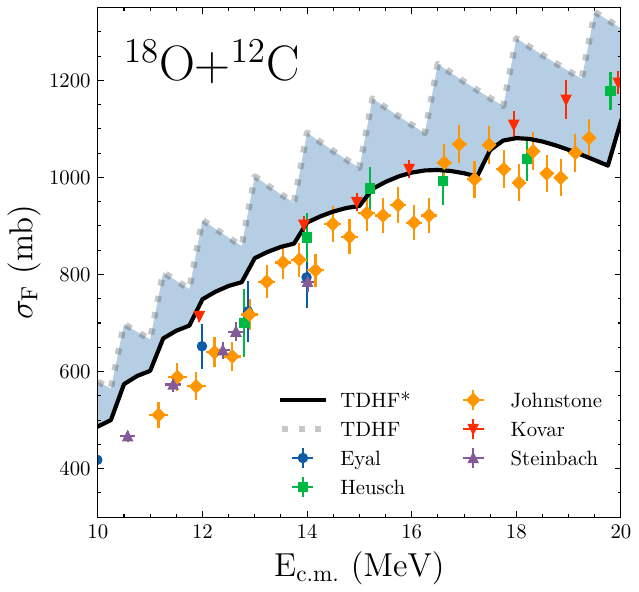}
\caption{ 
Similar as in Fig.\,\ref{fig:Fig2} but for
the $^{18}$O+ $^{12}$C reaction. Experimental data are taken from Refs. \cite{Eyal76} (blue circles), \cite {Heusch82} (green squares), \cite{Johnstone22} (orange diamonds) , \cite{Kovar79} (red upside-down triangles), and \cite{Steinbach14a} (purple triangles). 
}
\label{fig:Fig4}
\end{figure}

A more detailed comparison of the measured and calculated 
fusion excitation functions is provided in Figs.~\ref{fig:Fig2}-\ref{fig:Fig4}.
We first discuss the $^{16}$O+$^{12}$C reaction as it provides an excellent reference due to the rigid nature of the $^{16}$O projectile. 
As shown in Fig.~\ref{fig:Fig2},
for $\Ecm< 14$\,MeV, the   TDHF$^*$ method  provides a good description of the fusion excitation function due to the addition of successive  $\ell$-waves. For $\Ecm>11$\,MeV,  TDHF$^*$  systematically 
overestimates the measured excitation function,  although the oscillating behavior of the cross section is well reproduced.
The raw TDHF method  systematically overshoots the data.

Overestimation of the fusion cross-section at higher energies by TDHF has typically been attributed to the existence of breakup channels in the experimental data that are not properly represented in TDHF, though the full extent of this effect is an open question. Our TDHF$^*$ calculations indicate that a more accurate  description of transmission probabilities reduces the need for invoking breakup channels.
All in all, the description of the reference reaction $^{16}$O+ $^{12}$C by the parameter-free TDHF$^*$ approach is satisfactory.

Having established the success of  TDHF$^*$ in describing the  $^{16}$O + $^{12}$C reaction, we investigate the impact on fusion introduced by the addition of a single neutron to the projectile. 
Figure~\ref{fig:Fig3} illustrates the case of $^{17}$O + $^{12}$C.
The experimental data were collected in recent active thick-target measurements \cite{Asher21a,Hudan23} along with earlier thin-target measurements \cite{Tighe93, Hertz78, Eyal76}. 
It is to be noted that the close examination of  different experimental
datasets for $^{17}$O reveals some significant differences. For  $\Ecm\sim 14$\,MeV the data of \cite{Asher21a} and the lowest energy point from \cite{Hertz78} suggest a pronounced dip in the cross section differing from the data of \cite{Hudan23, Eyal76}. The accuracy of the thick-target data in Ref.\,\cite{Hudan23} has been corroborated by comparing the measured cross-section with thin-target measurement of the fusion cross-section of mirror nuclei. The magnitude of the dip at 
$\Ecm \sim 14$
\,MeV is significantly reduced as compared to \cite{Asher21a} and is shifted to slightly higher energy. 
Also, at the lowest energies shown, the data of Ref.~\cite{Tighe93} appears slightly low relative to the data from both  \cite{Hudan23} and \cite{Eyal76} which are in a reasonable agreement. As the data of Ref.\,\cite{Hudan23}  are self-normalizing, in our opinion, they provide a more accurate measure of the fusion cross section. 

The deviation from smooth behavior of the excitation function evident for the case of 
$^{16}$O + $^{12}$C, is also apparent in the case of the $^{17}$O but
the pronounced zigzag pattern  in the cross-section, as seen in the $^{16}$O data, is harder to quantify. 
The TDHF* calculations for this reaction significantly overestimate the measured cross section for $14<\Ecm<21$\,MeV. There are several possible reasons for this, including  neutron transfer which does not lead to fusion.
The impact of transfer on the fusion probabilities was estimated by checking the isovector fusion potentials extracted from DC-TDHF in a similar procedure to Ref.~\cite{godbey2017}. As seen in Fig.~S1 of \cite{sup},
the magnitude of the isovector contribution for $^{17}$O is less than that of $^{18}$O, suggesting that any transfer effects at the mean-field level will not account for the significant suppression in above barrier cross sections seen in experiment.
The presence of nucleonic cluster-like structures in the 
transient configurations can be probed by TDHF, see, e.g., \cite{Schuetrumpf2017}. However, the TDHF results shown in Fig.~\ref{fig:Fig1} do not show any appreciable reduction of $\sigma_{\rm F}$ for $^{17}$O. On the contrary, the predicted cross section for $^{17}$O systematically exceeds the $^{16}$O ``reference''. 

Since the odd neutron in $^{17}$O occupies the $0d_{5/2}$ orbit leading to the 5/2$^+$ ground state of $^{17}$O, some increase of the fusion barrier may be possible due to  a hindrance factor of fusion by specialization energy  -- an increase in the barrier due to angular momentum conservation \cite{Hofmann1997}. This effect, considered for fission, has so far not been considered by theoretical approaches to heavy-ion fusion. In particular, it is not accounted for by TDHF which does not conserve angular momentum. An experimental argument against this scenario, however, is the similarity of the measured fusion excitation functions for $^{16}$O and $^{17}$O projectiles at low energies seen in Fig.~\ref{fig:Fig1}.

We now examine the impact of two  valence neutrons
in $^{18}$O.
The  excitation function for $^{18}$O + $^{12}$C shown in Fig.~\ref{fig:Fig4} utilizes thin-target measurements \cite{Eyal76,Heusch82,Steinbach14a} together with recent active thick-target data \cite{Johnstone22}. 
While the experimental data exhibit oscillations, the presence of sharp resonant-like structures is absent. The TDHF$^*$ model with pairing provides  a reasonably good overall agreement with the data although the  calculations slightly overestimate the  data.

Pairing correlations are expected to effectively increase the fusion barrier, hence decrease fusion cross sections \cite{Steinbach14a,Magierski2017}. The experimental data in Fig.~\ref{fig:Fig1} do not manifest such a reduction for $^{18}$O: actually its fusion excitation function exceeds that for $^{16}$O for $\Ecm>14$\,MeV. This result implies that the impact of pairing correlations on 
$\sigma_{\rm F}$ in $^{18}$O is minor, consistent with the similarity of predicted fusion excitation function for $^{18}$O with that of $ ^{16,17}$O in Fig.~\ref{fig:Fig1}.

{\it Summary.---} We have presented a framework for using a microscopic, parameter-free TDHF$^*$ model to investigate  fusion excitation functions in the oxygen isotopes. To obtain $\sigma_{\rm F}(E)$
with sufficient resolution, multiple experimental datasets were combined. The resulting data reveal oscillatory structures consistent with the presence of  different $\ell$-wave barriers. To accurately describe the experimental data,  an extension of the standard TDHF approach was required to calculate the fusion  penetrability  directly from the DC-TDHF potential. 
The resulting TDHF$^*$ model provided a reasonably good description for the reference case
 of the $^{16}$O-induced fusion, including the  reproduction of oscillatory structures. 
A slightly worse, but still acceptable agreement with experiment was obtained for the $^{18}$O-induced fusion. An appreciable reduction of the experimental fusion excitation function for $^{17}$O remains a puzzle.

Several possible explanations exist for the remaining discrepancies between experiment and theory:  the effect of breakup and transfer channels, an imperfect description of $\ell$-dependent fusion barriers by TDHF, or the presence of transient configurations involving nucleonic clusters. Distinguishing between these possibilities requires advances on both experimental and theoretical fronts. Systematic high-resolution, exclusive measurements of heavy-ion fusion  and transfer/breakup measurements along isotopic chains is necessary in order to establish the limits of breakup and transfer channels.
This experimental data, paired with continued investment in high-performance computing, will be critical in enabling the development of a more complete beyond-mean-field description of heavy-ion fusion.


\begin{acknowledgments}
This work was supported by the U.S. Department of Energy Office of Science under Grant Nos. 
DE-FG02-88ER-40404 (Indiana University), DOE-DE-SC0013365
and DE-SC0023175
(Michigan State University), and DOE-DE-NA0004074 (NNSA, the Stewardship Science Academic Alliances program).
This work was supported in part through computational resources and services provided by the Institute for Cyber-Enabled Research at Michigan State University. 

\end{acknowledgments}


\begin{thebibliography}{40}%
\makeatletter
\providecommand \@ifxundefined [1]{%
 \@ifx{#1\undefined}
}%
\providecommand \@ifnum [1]{%
 \ifnum #1\expandafter \@firstoftwo
 \else \expandafter \@secondoftwo
 \fi
}%
\providecommand \@ifx [1]{%
 \ifx #1\expandafter \@firstoftwo
 \else \expandafter \@secondoftwo
 \fi
}%
\providecommand \natexlab [1]{#1}%
\providecommand \enquote  [1]{``#1''}%
\providecommand \bibnamefont  [1]{#1}%
\providecommand \bibfnamefont [1]{#1}%
\providecommand \citenamefont [1]{#1}%
\providecommand \href@noop [0]{\@secondoftwo}%
\providecommand \href [0]{\begingroup \@sanitize@url \@href}%
\providecommand \@href[1]{\@@startlink{#1}\@@href}%
\providecommand \@@href[1]{\endgroup#1\@@endlink}%
\providecommand \@sanitize@url [0]{\catcode `\\12\catcode `\$12\catcode
  `\&12\catcode `\#12\catcode `\^12\catcode `\_12\catcode `\%12\relax}%
\providecommand \@@startlink[1]{}%
\providecommand \@@endlink[0]{}%
\providecommand \url  [0]{\begingroup\@sanitize@url \@url }%
\providecommand \@url [1]{\endgroup\@href {#1}{\urlprefix }}%
\providecommand \urlprefix  [0]{URL }%
\providecommand \Eprint [0]{\href }%
\providecommand \doibase [0]{https://doi.org/}%
\providecommand \selectlanguage [0]{\@gobble}%
\providecommand \bibinfo  [0]{\@secondoftwo}%
\providecommand \bibfield  [0]{\@secondoftwo}%
\providecommand \translation [1]{[#1]}%
\providecommand \BibitemOpen [0]{}%
\providecommand \bibitemStop [0]{}%
\providecommand \bibitemNoStop [0]{.\EOS\space}%
\providecommand \EOS [0]{\spacefactor3000\relax}%
\providecommand \BibitemShut  [1]{\csname bibitem#1\endcsname}%
\let\auto@bib@innerbib\@empty
\bibitem [{\citenamefont {Bromley}\ \emph {et~al.}(1960)\citenamefont
  {Bromley}, \citenamefont {Kuehner},\ and\ \citenamefont
  {Almquist}}]{Bromley60}%
  \BibitemOpen
  \bibfield  {author} {\bibinfo {author} {\bibfnamefont {D.}~\bibnamefont
  {Bromley}}, \bibinfo {author} {\bibfnamefont {J.}~\bibnamefont {Kuehner}},\
  and\ \bibinfo {author} {\bibfnamefont {E.}~\bibnamefont {Almquist}},\
  }\bibfield  {title} {\bibinfo {title} {Resonant elastic scattering of
  $^{12}${C} by carbon},\ }\href {https://doi.org/10.1103/PhysRevLett.4.365}
  {\bibfield  {journal} {\bibinfo  {journal} {Phys. Rev. Lett.}\ }\textbf
  {\bibinfo {volume} {4}},\ \bibinfo {pages} {365} (\bibinfo {year}
  {1960})}\BibitemShut {NoStop}%
\bibitem [{\citenamefont {Vogt}\ and\ \citenamefont {McManus}(1960)}]{Vogt60}%
  \BibitemOpen
  \bibfield  {author} {\bibinfo {author} {\bibfnamefont {E.}~\bibnamefont
  {Vogt}}\ and\ \bibinfo {author} {\bibfnamefont {H.}~\bibnamefont {McManus}},\
  }\bibfield  {title} {\bibinfo {title} {"molecular" states formed by two
  carbon nuclei},\ }\href {https://doi.org/10.1103/PhysRevLett.4.518}
  {\bibfield  {journal} {\bibinfo  {journal} {Phys. Rev. Lett.}\ }\textbf
  {\bibinfo {volume} {4}},\ \bibinfo {pages} {518} (\bibinfo {year}
  {1960})}\BibitemShut {NoStop}%
\bibitem [{\citenamefont {Sperr}\ \emph
  {et~al.}(1976{\natexlab{a}})\citenamefont {Sperr}, \citenamefont {Braid},
  \citenamefont {Y.~Eisen}, \citenamefont {Kovar}, \citenamefont {Prosser},
  \citenamefont {Schiffer}, \citenamefont {Tabor},\ and\ \citenamefont
  {Vigdor}}]{Sperr76}%
  \BibitemOpen
  \bibfield  {author} {\bibinfo {author} {\bibfnamefont {P.}~\bibnamefont
  {Sperr}}, \bibinfo {author} {\bibfnamefont {T.~H.}\ \bibnamefont {Braid}},
  \bibinfo {author} {\bibfnamefont {Y.}~\bibnamefont {Y.~Eisen}}, \bibinfo
  {author} {\bibfnamefont {D.~G.}\ \bibnamefont {Kovar}}, \bibinfo {author}
  {\bibfnamefont {F.~W.}\ \bibnamefont {Prosser}}, \bibinfo {author}
  {\bibfnamefont {J.~P.}\ \bibnamefont {Schiffer}}, \bibinfo {author}
  {\bibfnamefont {S.~L.}\ \bibnamefont {Tabor}},\ and\ \bibinfo {author}
  {\bibfnamefont {S.}~\bibnamefont {Vigdor}},\ }\bibfield  {title} {\bibinfo
  {title} {Fusion cross sections of light heavy-ion systems: Resonances and
  shell effects},\ }\href {https://doi.org/10.1103/PhysRevLett.37.321}
  {\bibfield  {journal} {\bibinfo  {journal} {Phys. Rev. Lett.}\ }\textbf
  {\bibinfo {volume} {37}},\ \bibinfo {pages} {321} (\bibinfo {year}
  {1976}{\natexlab{a}})}\BibitemShut {NoStop}%
\bibitem [{\citenamefont {Sperr}\ \emph
  {et~al.}(1976{\natexlab{b}})\citenamefont {Sperr}, \citenamefont {Vigdor},
  \citenamefont {Eisen}, \citenamefont {Henning}, \citenamefont {Kovar},
  \citenamefont {Ophel},\ and\ \citenamefont {Zeidman}}]{Sperr76a}%
  \BibitemOpen
  \bibfield  {author} {\bibinfo {author} {\bibfnamefont {P.}~\bibnamefont
  {Sperr}}, \bibinfo {author} {\bibfnamefont {S.}~\bibnamefont {Vigdor}},
  \bibinfo {author} {\bibfnamefont {Y.}~\bibnamefont {Eisen}}, \bibinfo
  {author} {\bibfnamefont {W.}~\bibnamefont {Henning}}, \bibinfo {author}
  {\bibfnamefont {D.~G.}\ \bibnamefont {Kovar}}, \bibinfo {author}
  {\bibfnamefont {T.}~\bibnamefont {Ophel}},\ and\ \bibinfo {author}
  {\bibfnamefont {B.}~\bibnamefont {Zeidman}},\ }\bibfield  {title} {\bibinfo
  {title} {Oscillations in the excitation function for complete fusion of
  $^{16}${O}+$^{12}${C}},\ }\href {https://doi.org/10.1103/PhysRevLett.36.405}
  {\bibfield  {journal} {\bibinfo  {journal} {Phys. Rev. Lett.}\ }\textbf
  {\bibinfo {volume} {36}},\ \bibinfo {pages} {405} (\bibinfo {year}
  {1976}{\natexlab{b}})}\BibitemShut {NoStop}%
\bibitem [{\citenamefont {Kovar}\ \emph {et~al.}(1979)\citenamefont {Kovar}
  \emph {et~al.}}]{Kovar79}%
  \BibitemOpen
  \bibfield  {author} {\bibinfo {author} {\bibfnamefont {D.~G.}\ \bibnamefont
  {Kovar}} \emph {et~al.},\ }\bibfield  {title} {\bibinfo {title} {Systematics
  of carbon- and oxygen-induced fusion on nuclei with $12\le${A}$\le{}19$},\
  }\href {https://doi.org/10.1103/PhysRevC.20.1305} {\bibfield  {journal}
  {\bibinfo  {journal} {Phys. Rev. C}\ }\textbf {\bibinfo {volume} {20}},\
  \bibinfo {pages} {1305} (\bibinfo {year} {1979})}\BibitemShut {NoStop}%
\bibitem [{\citenamefont {Fernandez}\ \emph {et~al.}(1978)\citenamefont
  {Fernandez}, \citenamefont {Gaarde}, \citenamefont {Larsen}, \citenamefont
  {Pontoppidan},\ and\ \citenamefont {Videbaek}}]{Fernandez78}%
  \BibitemOpen
  \bibfield  {author} {\bibinfo {author} {\bibfnamefont {B.}~\bibnamefont
  {Fernandez}}, \bibinfo {author} {\bibfnamefont {C.}~\bibnamefont {Gaarde}},
  \bibinfo {author} {\bibfnamefont {J.}~\bibnamefont {Larsen}}, \bibinfo
  {author} {\bibfnamefont {S.}~\bibnamefont {Pontoppidan}},\ and\ \bibinfo
  {author} {\bibfnamefont {F.}~\bibnamefont {Videbaek}},\ }\bibfield  {title}
  {\bibinfo {title} {{F}usion cross sections for the $^{16}${O}+$^{16}${O}
  reaction},\ }\href {https://doi.org/10.1016/j.nuclphysa.1978.02.018}
  {\bibfield  {journal} {\bibinfo  {journal} {Nucl. Phys. A}\ }\textbf
  {\bibinfo {volume} {306}},\ \bibinfo {pages} {259} (\bibinfo {year}
  {1978})}\BibitemShut {NoStop}%
\bibitem [{\citenamefont {Kolata}\ \emph {et~al.}(1979)\citenamefont {Kolata},
  \citenamefont {Freeman}, \citenamefont {Haas}, \citenamefont {Heusch},\ and\
  \citenamefont {Gallmann}}]{Kolata79}%
  \BibitemOpen
  \bibfield  {author} {\bibinfo {author} {\bibfnamefont {J.}~\bibnamefont
  {Kolata}}, \bibinfo {author} {\bibfnamefont {R.}~\bibnamefont {Freeman}},
  \bibinfo {author} {\bibfnamefont {F.}~\bibnamefont {Haas}}, \bibinfo {author}
  {\bibfnamefont {B.}~\bibnamefont {Heusch}},\ and\ \bibinfo {author}
  {\bibfnamefont {A.}~\bibnamefont {Gallmann}},\ }\bibfield  {title} {\bibinfo
  {title} {Gross and intermediate-width structure in the interaction of
  $^{16}${O} with $^{16}${O}},\ }\href
  {https://doi.org/10.1103/PhysRevC.19.2237} {\bibfield  {journal} {\bibinfo
  {journal} {Phys. Rev. C}\ }\textbf {\bibinfo {volume} {19}},\ \bibinfo
  {pages} {2237} (\bibinfo {year} {1979})}\BibitemShut {NoStop}%
\bibitem [{\citenamefont {Tserruya}\ \emph {et~al.}(1978)\citenamefont
  {Tserruya}, \citenamefont {Risen}, \citenamefont {Pelte}, \citenamefont
  {Gavron}, \citenamefont {OeschIer}, \citenamefont {Berndt},\ and\
  \citenamefont {Harney}}]{Tserruya78}%
  \BibitemOpen
  \bibfield  {author} {\bibinfo {author} {\bibfnamefont {I.}~\bibnamefont
  {Tserruya}}, \bibinfo {author} {\bibfnamefont {Y.}~\bibnamefont {Risen}},
  \bibinfo {author} {\bibfnamefont {D.}~\bibnamefont {Pelte}}, \bibinfo
  {author} {\bibfnamefont {A.}~\bibnamefont {Gavron}}, \bibinfo {author}
  {\bibfnamefont {H.}~\bibnamefont {OeschIer}}, \bibinfo {author}
  {\bibfnamefont {D.}~\bibnamefont {Berndt}},\ and\ \bibinfo {author}
  {\bibfnamefont {H.}~\bibnamefont {Harney}},\ }\bibfield  {title} {\bibinfo
  {title} {Total fusion cross section for the $^{16}${O}+ $^{16}${O} system},\
  }\href {https://doi.org/10.1103/PhysRevC.18.1688} {\bibfield  {journal}
  {\bibinfo  {journal} {Phys. Rev. C}\ }\textbf {\bibinfo {volume} {18}},\
  \bibinfo {pages} {1688} (\bibinfo {year} {1978})}\BibitemShut {NoStop}%
\bibitem [{\citenamefont {Esbensen}(2012)}]{Esbensen12}%
  \BibitemOpen
  \bibfield  {author} {\bibinfo {author} {\bibfnamefont {H.}~\bibnamefont
  {Esbensen}},\ }\bibfield  {title} {\bibinfo {title} {Structures in
  high-energy fusion data},\ }\href
  {https://doi.org/10.1103/PhysRevC.85.064611} {\bibfield  {journal} {\bibinfo
  {journal} {Phys. Rev. C}\ }\textbf {\bibinfo {volume} {85}},\ \bibinfo
  {pages} {064611} (\bibinfo {year} {2012})}\BibitemShut {NoStop}%
\bibitem [{\citenamefont {Wong}(2012)}]{Wong12}%
  \BibitemOpen
  \bibfield  {author} {\bibinfo {author} {\bibfnamefont {C.~Y.}\ \bibnamefont
  {Wong}},\ }\bibfield  {title} {\bibinfo {title} {Reaction cross sections in
  heavy-ion collisions},\ }\href {https://doi.org/10.1103/PhysRevC.86.064603}
  {\bibfield  {journal} {\bibinfo  {journal} {Phys. Rev. C}\ }\textbf {\bibinfo
  {volume} {86}},\ \bibinfo {pages} {064603} (\bibinfo {year}
  {2012})}\BibitemShut {NoStop}%
\bibitem [{\citenamefont {Simenel}\ \emph
  {et~al.}(2013{\natexlab{a}})\citenamefont {Simenel}, \citenamefont {Keser},
  \citenamefont {Umar},\ and\ \citenamefont {Oberacker}}]{Simenel13}%
  \BibitemOpen
  \bibfield  {author} {\bibinfo {author} {\bibfnamefont {C.}~\bibnamefont
  {Simenel}}, \bibinfo {author} {\bibfnamefont {R.}~\bibnamefont {Keser}},
  \bibinfo {author} {\bibfnamefont {A.}~\bibnamefont {Umar}},\ and\ \bibinfo
  {author} {\bibfnamefont {V.}~\bibnamefont {Oberacker}},\ }\bibfield  {title}
  {\bibinfo {title} {Microscopic study of $^{16}${O}+$^{16}${O} fusion},\
  }\href {https://doi.org/10.1103/PhysRevC.88.024617} {\bibfield  {journal}
  {\bibinfo  {journal} {Phys. Rev. C}\ }\textbf {\bibinfo {volume} {88}},\
  \bibinfo {pages} {024617} (\bibinfo {year} {2013}{\natexlab{a}})}\BibitemShut
  {NoStop}%
\bibitem [{\citenamefont {Rowley}\ and\ \citenamefont
  {Hagino}(2015)}]{Rowley15}%
  \BibitemOpen
  \bibfield  {author} {\bibinfo {author} {\bibfnamefont {N.}~\bibnamefont
  {Rowley}}\ and\ \bibinfo {author} {\bibfnamefont {K.}~\bibnamefont
  {Hagino}},\ }\bibfield  {title} {\bibinfo {title} {Examination of fusion
  cross sections and fusion oscillations with a generalized {W}ong formula},\
  }\href {https://doi.org/http://dx.doi.org/10.1103/PhysRevC.91.044617}
  {\bibfield  {journal} {\bibinfo  {journal} {Phys. Rev. C}\ }\textbf {\bibinfo
  {volume} {91}},\ \bibinfo {pages} {044617} (\bibinfo {year}
  {2015})}\BibitemShut {NoStop}%
\bibitem [{\citenamefont {Frawley}\ \emph {et~al.}(1982)\citenamefont
  {Frawley}, \citenamefont {Fletcher},\ and\ \citenamefont
  {Dennis}}]{Frawley82}%
  \BibitemOpen
  \bibfield  {author} {\bibinfo {author} {\bibfnamefont {A.~D.}\ \bibnamefont
  {Frawley}}, \bibinfo {author} {\bibfnamefont {N.~R.}\ \bibnamefont
  {Fletcher}},\ and\ \bibinfo {author} {\bibfnamefont {L.~C.}\ \bibnamefont
  {Dennis}},\ }\bibfield  {title} {\bibinfo {title} {Resonances in the
  $^{16}${O} + $^{12}${C} fusion cross section between {E}$_{c.m.}$ $=$ 12 and
  20 {M}e{V}},\ }\href {https://doi.org/10.1103/PhysRevC.25.860} {\bibfield
  {journal} {\bibinfo  {journal} {Phys. Rev. C}\ }\textbf {\bibinfo {volume}
  {25}},\ \bibinfo {pages} {860} (\bibinfo {year} {1982})}\BibitemShut
  {NoStop}%
\bibitem [{\citenamefont {Hudan}\ \emph {et~al.}(2023)\citenamefont {Hudan},
  \citenamefont {Johnstone}, \citenamefont {Kumar}, \citenamefont {deSouza},
  \citenamefont {Allen}, \citenamefont {Bardayan}, \citenamefont {Blankstein},
  \citenamefont {Boomershine}, \citenamefont {Carmichael}, \citenamefont
  {Clark}, \citenamefont {Coil}, \citenamefont {Henderson}, \citenamefont
  {O’Malley},\ and\ \citenamefont {von Seeger}}]{Hudan23}%
  \BibitemOpen
  \bibfield  {author} {\bibinfo {author} {\bibfnamefont {S.}~\bibnamefont
  {Hudan}}, \bibinfo {author} {\bibfnamefont {J.~E.}\ \bibnamefont
  {Johnstone}}, \bibinfo {author} {\bibfnamefont {R.}~\bibnamefont {Kumar}},
  \bibinfo {author} {\bibfnamefont {R.~T.}\ \bibnamefont {deSouza}}, \bibinfo
  {author} {\bibfnamefont {J.}~\bibnamefont {Allen}}, \bibinfo {author}
  {\bibfnamefont {D.~W.}\ \bibnamefont {Bardayan}}, \bibinfo {author}
  {\bibfnamefont {D.}~\bibnamefont {Blankstein}}, \bibinfo {author}
  {\bibfnamefont {C.}~\bibnamefont {Boomershine}}, \bibinfo {author}
  {\bibfnamefont {S.}~\bibnamefont {Carmichael}}, \bibinfo {author}
  {\bibfnamefont {A.}~\bibnamefont {Clark}}, \bibinfo {author} {\bibfnamefont
  {S.}~\bibnamefont {Coil}}, \bibinfo {author} {\bibfnamefont {S.~L.}\
  \bibnamefont {Henderson}}, \bibinfo {author} {\bibfnamefont {P.~D.}\
  \bibnamefont {O’Malley}},\ and\ \bibinfo {author} {\bibfnamefont {W.~W.}\
  \bibnamefont {von Seeger}},\ }\bibfield  {title} {\bibinfo {title}
  {Quantifying resonance behavior in the fusion of $^{17}${O} with $^{12}${C}
  at above-barrier energies},\ }\href
  {https://doi.org/10.1103/PhysRevC.107.064612} {\bibfield  {journal} {\bibinfo
   {journal} {Phys. Rev. C}\ }\textbf {\bibinfo {volume} {107}},\ \bibinfo
  {pages} {064612} (\bibinfo {year} {2023})}\BibitemShut {NoStop}%
\bibitem [{\citenamefont {Johnstone}\ \emph {et~al.}(2021)\citenamefont
  {Johnstone} \emph {et~al.}}]{Johnstone22}%
  \BibitemOpen
  \bibfield  {author} {\bibinfo {author} {\bibfnamefont {J.}~\bibnamefont
  {Johnstone}} \emph {et~al.},\ }\bibfield  {title} {\bibinfo {title}
  {Improving the characterization of fusion in a {M}u{SIC} detector by spatial
  localization},\ }\href {https://doi.org/10.1016/j.nima.2022.166212}
  {\bibfield  {journal} {\bibinfo  {journal} {Nucl. Instr. Meth. A}\ }\textbf
  {\bibinfo {volume} {1025}},\ \bibinfo {pages} {166212} (\bibinfo {year}
  {2021})}\BibitemShut {NoStop}%
\bibitem [{\citenamefont {Steinbach}\ \emph {et~al.}(2014)\citenamefont
  {Steinbach}, \citenamefont {Vadas}, \citenamefont {Schmidt}, \citenamefont
  {Haycraft}, \citenamefont {Hudan}, \citenamefont {deSouza}, \citenamefont
  {Baby}, \citenamefont {Kuvin}, \citenamefont {Wiedenh\"over}, \citenamefont
  {Umar},\ and\ \citenamefont {Oberacker}}]{Steinbach14a}%
  \BibitemOpen
  \bibfield  {author} {\bibinfo {author} {\bibfnamefont {T.~K.}\ \bibnamefont
  {Steinbach}}, \bibinfo {author} {\bibfnamefont {J.}~\bibnamefont {Vadas}},
  \bibinfo {author} {\bibfnamefont {J.}~\bibnamefont {Schmidt}}, \bibinfo
  {author} {\bibfnamefont {C.}~\bibnamefont {Haycraft}}, \bibinfo {author}
  {\bibfnamefont {S.}~\bibnamefont {Hudan}}, \bibinfo {author} {\bibfnamefont
  {R.~T.}\ \bibnamefont {deSouza}}, \bibinfo {author} {\bibfnamefont {L.~T.}\
  \bibnamefont {Baby}}, \bibinfo {author} {\bibfnamefont {S.~A.}\ \bibnamefont
  {Kuvin}}, \bibinfo {author} {\bibfnamefont {I.}~\bibnamefont
  {Wiedenh\"over}}, \bibinfo {author} {\bibfnamefont {A.~S.}\ \bibnamefont
  {Umar}},\ and\ \bibinfo {author} {\bibfnamefont {V.~E.}\ \bibnamefont
  {Oberacker}},\ }\bibfield  {title} {\bibinfo {title} {Sub-barrier enhancement
  of fusion as compared to a microscopic method in
  $^{18}\mathrm{O}+{}^{12}\mathrm{C}$},\ }\href
  {https://doi.org/10.1103/PhysRevC.90.041603} {\bibfield  {journal} {\bibinfo
  {journal} {Phys. Rev. C}\ }\textbf {\bibinfo {volume} {90}},\ \bibinfo
  {pages} {041603(R)} (\bibinfo {year} {2014})}\BibitemShut {NoStop}%
\bibitem [{\citenamefont {Gollan}\ \emph {et~al.}(2021)\citenamefont {Gollan},
  \citenamefont {Abriola}, \citenamefont {Arazi}, \citenamefont {Cardona},
  \citenamefont {de~Barbara}, \citenamefont {J.~de Jesus}, \citenamefont
  {Betan}, \citenamefont {Lubian}, \citenamefont {Pacheco}, \citenamefont
  {Paes}, \citenamefont {Schneider},\ and\ \citenamefont {Soler}}]{Gollan21}%
  \BibitemOpen
  \bibfield  {author} {\bibinfo {author} {\bibfnamefont {F.}~\bibnamefont
  {Gollan}}, \bibinfo {author} {\bibfnamefont {D.}~\bibnamefont {Abriola}},
  \bibinfo {author} {\bibfnamefont {A.}~\bibnamefont {Arazi}}, \bibinfo
  {author} {\bibfnamefont {M.~A.}\ \bibnamefont {Cardona}}, \bibinfo {author}
  {\bibfnamefont {E.}~\bibnamefont {de~Barbara}}, \bibinfo {author}
  {\bibfnamefont {D.~H.}\ \bibnamefont {J.~de Jesus}}, \bibinfo {author}
  {\bibfnamefont {R.~M.~I.}\ \bibnamefont {Betan}}, \bibinfo {author}
  {\bibfnamefont {J.}~\bibnamefont {Lubian}}, \bibinfo {author} {\bibfnamefont
  {A.~J.}\ \bibnamefont {Pacheco}}, \bibinfo {author} {\bibfnamefont
  {B.}~\bibnamefont {Paes}}, \bibinfo {author} {\bibfnamefont {D.}~\bibnamefont
  {Schneider}},\ and\ \bibinfo {author} {\bibfnamefont {H.~O.}\ \bibnamefont
  {Soler}},\ }\bibfield  {title} {\bibinfo {title} {One-neutron transfer,
  complete fusion, and incomplete fusion from the $^9${B}e + $^{197}${A}u
  reaction},\ }\href {https://doi.org/10.1103/PhysRevC.104.024609} {\bibfield
  {journal} {\bibinfo  {journal} {Phys. Rev. C}\ }\textbf {\bibinfo {volume}
  {104}},\ \bibinfo {pages} {024609} (\bibinfo {year} {2021})}\BibitemShut
  {NoStop}%
\bibitem [{\citenamefont {Eyal}\ \emph {et~al.}(1976)\citenamefont {Eyal},
  \citenamefont {Beckerman}, \citenamefont {Chechik}, \citenamefont
  {Fraenkel},\ and\ \citenamefont {Stocker}}]{Eyal76}%
  \BibitemOpen
  \bibfield  {author} {\bibinfo {author} {\bibfnamefont {Y.}~\bibnamefont
  {Eyal}}, \bibinfo {author} {\bibfnamefont {M.}~\bibnamefont {Beckerman}},
  \bibinfo {author} {\bibfnamefont {R.}~\bibnamefont {Chechik}}, \bibinfo
  {author} {\bibfnamefont {Z.}~\bibnamefont {Fraenkel}},\ and\ \bibinfo
  {author} {\bibfnamefont {H.}~\bibnamefont {Stocker}},\ }\bibfield  {title}
  {\bibinfo {title} {{N}uclear size and boundary effects on fusion barrier of
  oxygen with carbon},\ }\href {https://doi.org/10.1103/PhysRevC.13.1527}
  {\bibfield  {journal} {\bibinfo  {journal} {Phys. Rev. C}\ }\textbf {\bibinfo
  {volume} {13}},\ \bibinfo {pages} {1527} (\bibinfo {year}
  {1976})}\BibitemShut {NoStop}%
\bibitem [{\citenamefont {Cujec}\ and\ \citenamefont {Barnes}(1976)}]{Cujec76}%
  \BibitemOpen
  \bibfield  {author} {\bibinfo {author} {\bibfnamefont {B.}~\bibnamefont
  {Cujec}}\ and\ \bibinfo {author} {\bibfnamefont {C.}~\bibnamefont {Barnes}},\
  }\bibfield  {title} {\bibinfo {title} {Total reaction cross section for
  $^{12}${C} + $^{16}${O} below the coulomb barrier},\ }\href@noop {}
  {\bibfield  {journal} {\bibinfo  {journal} {Nucl. Phys. A}\ }\textbf
  {\bibinfo {volume} {266}},\ \bibinfo {pages} {461} (\bibinfo {year}
  {1976})}\BibitemShut {NoStop}%
\bibitem [{\citenamefont {Godbey}\ \emph {et~al.}(2017)\citenamefont {Godbey},
  \citenamefont {Umar},\ and\ \citenamefont {Simenel}}]{godbey2017}%
  \BibitemOpen
  \bibfield  {author} {\bibinfo {author} {\bibfnamefont {K.}~\bibnamefont
  {Godbey}}, \bibinfo {author} {\bibfnamefont {A.~S.}\ \bibnamefont {Umar}},\
  and\ \bibinfo {author} {\bibfnamefont {C.}~\bibnamefont {Simenel}},\
  }\bibfield  {title} {\bibinfo {title} {Dependence of fusion on isospin
  dynamics},\ }\href {https://doi.org/10.1103/PhysRevC.95.011601} {\bibfield
  {journal} {\bibinfo  {journal} {Phys. Rev. C}\ }\textbf {\bibinfo {volume}
  {95}},\ \bibinfo {pages} {011601} (\bibinfo {year} {2017})}\BibitemShut
  {NoStop}%
\bibitem [{\citenamefont {Simenel}\ \emph {et~al.}(2020)\citenamefont
  {Simenel}, \citenamefont {Godbey},\ and\ \citenamefont {Umar}}]{simenel2020}%
  \BibitemOpen
  \bibfield  {author} {\bibinfo {author} {\bibfnamefont {C.}~\bibnamefont
  {Simenel}}, \bibinfo {author} {\bibfnamefont {K.}~\bibnamefont {Godbey}},\
  and\ \bibinfo {author} {\bibfnamefont {A.~S.}\ \bibnamefont {Umar}},\
  }\bibfield  {title} {\bibinfo {title} {Timescales of quantum equilibration,
  dissipation and fluctuation in nuclear collisions},\ }\href
  {https://doi.org/10.1103/PhysRevLett.124.212504} {\bibfield  {journal}
  {\bibinfo  {journal} {Phys. Rev. Lett.}\ }\textbf {\bibinfo {volume} {124}},\
  \bibinfo {pages} {212504} (\bibinfo {year} {2020})}\BibitemShut {NoStop}%
\bibitem [{\citenamefont {Simenel}\ \emph {et~al.}(2017)\citenamefont
  {Simenel}, \citenamefont {Umar}, \citenamefont {Godbey}, \citenamefont
  {Dasgupta},\ and\ \citenamefont {Hinde}}]{simenel2017}%
  \BibitemOpen
  \bibfield  {author} {\bibinfo {author} {\bibfnamefont {C.}~\bibnamefont
  {Simenel}}, \bibinfo {author} {\bibfnamefont {A.~S.}\ \bibnamefont {Umar}},
  \bibinfo {author} {\bibfnamefont {K.}~\bibnamefont {Godbey}}, \bibinfo
  {author} {\bibfnamefont {M.}~\bibnamefont {Dasgupta}},\ and\ \bibinfo
  {author} {\bibfnamefont {D.~J.}\ \bibnamefont {Hinde}},\ }\bibfield  {title}
  {\bibinfo {title} {How the pauli exclusion principle affects fusion of atomic
  nuclei},\ }\href {https://doi.org/10.1103/PhysRevC.95.031601} {\bibfield
  {journal} {\bibinfo  {journal} {Phys. Rev. C}\ }\textbf {\bibinfo {volume}
  {95}},\ \bibinfo {pages} {031601} (\bibinfo {year} {2017})}\BibitemShut
  {NoStop}%
\bibitem [{\citenamefont {R.~Keser}\ \emph {et~al.}(2012)\citenamefont
  {R.~Keser}, \citenamefont {Umar},\ and\ \citenamefont {Oberacker}}]{Keser12}%
  \BibitemOpen
  \bibfield  {author} {\bibinfo {author} {\bibfnamefont {R.}~\bibnamefont
  {R.~Keser}}, \bibinfo {author} {\bibfnamefont {A.}~\bibnamefont {Umar}},\
  and\ \bibinfo {author} {\bibfnamefont {V.}~\bibnamefont {Oberacker}},\
  }\bibfield  {title} {\bibinfo {title} {Microscopic study of {C}a + {C}a
  fusion},\ }\href {https://doi.org/10.1103/PhysRevC.85.044606} {\bibfield
  {journal} {\bibinfo  {journal} {Phys. Rev. C}\ }\textbf {\bibinfo {volume}
  {85}},\ \bibinfo {pages} {044606} (\bibinfo {year} {2012})}\BibitemShut
  {NoStop}%
\bibitem [{\citenamefont {Simenel}\ \emph
  {et~al.}(2013{\natexlab{b}})\citenamefont {Simenel}, \citenamefont {Keser},
  \citenamefont {Umar},\ and\ \citenamefont {Oberacker}}]{simenel2013a}%
  \BibitemOpen
  \bibfield  {author} {\bibinfo {author} {\bibfnamefont {C.}~\bibnamefont
  {Simenel}}, \bibinfo {author} {\bibfnamefont {R.}~\bibnamefont {Keser}},
  \bibinfo {author} {\bibfnamefont {A.~S.}\ \bibnamefont {Umar}},\ and\
  \bibinfo {author} {\bibfnamefont {V.~E.}\ \bibnamefont {Oberacker}},\
  }\bibfield  {title} {\bibinfo {title} {Microscopic study of
  ${}^{16}\mathrm{O}+{}^{16}\mathrm{O}$ fusion},\ }\href
  {https://doi.org/10.1103/PhysRevC.88.024617} {\bibfield  {journal} {\bibinfo
  {journal} {Phys. Rev. C}\ }\textbf {\bibinfo {volume} {88}},\ \bibinfo
  {pages} {024617} (\bibinfo {year} {2013}{\natexlab{b}})}\BibitemShut
  {NoStop}%
\bibitem [{\citenamefont {Godbey}\ \emph {et~al.}(2019)\citenamefont {Godbey},
  \citenamefont {Simenel},\ and\ \citenamefont {Umar}}]{godbey2019b}%
  \BibitemOpen
  \bibfield  {author} {\bibinfo {author} {\bibfnamefont {K.}~\bibnamefont
  {Godbey}}, \bibinfo {author} {\bibfnamefont {C.}~\bibnamefont {Simenel}},\
  and\ \bibinfo {author} {\bibfnamefont {A.~S.}\ \bibnamefont {Umar}},\
  }\bibfield  {title} {\bibinfo {title} {Absence of hindrance in microscopic
  ${}^{12}\mathrm{C}+{}^{12}\mathrm{C}$ fusion study},\ }\href
  {https://doi.org/10.1103/PhysRevC.100.024619} {\bibfield  {journal} {\bibinfo
   {journal} {Phys. Rev. C}\ }\textbf {\bibinfo {volume} {100}},\ \bibinfo
  {pages} {024619} (\bibinfo {year} {2019})}\BibitemShut {NoStop}%
\bibitem [{\citenamefont {Umar}\ \emph {et~al.}(2012)\citenamefont {Umar},
  \citenamefont {Oberacker},\ and\ \citenamefont {Horowitz}}]{Umar12}%
  \BibitemOpen
  \bibfield  {author} {\bibinfo {author} {\bibfnamefont {A.~S.}\ \bibnamefont
  {Umar}}, \bibinfo {author} {\bibfnamefont {V.~E.}\ \bibnamefont
  {Oberacker}},\ and\ \bibinfo {author} {\bibfnamefont {C.~J.}\ \bibnamefont
  {Horowitz}},\ }\bibfield  {title} {\bibinfo {title} {{M}icroscopic
  sub-barrier fusion calculations for the neutron star crust},\ }\href
  {https://doi.org/10.1103/PhysRevC.85.055801} {\bibfield  {journal} {\bibinfo
  {journal} {Phys. Rev. C}\ }\textbf {\bibinfo {volume} {85}},\ \bibinfo
  {pages} {055801} (\bibinfo {year} {2012})}\BibitemShut {NoStop}%
\bibitem [{\citenamefont {deSouza}\ \emph {et~al.}(2013)\citenamefont
  {deSouza}, \citenamefont {Hudan}, \citenamefont {Oberacker},\ and\
  \citenamefont {Umar}}]{deSouza13}%
  \BibitemOpen
  \bibfield  {author} {\bibinfo {author} {\bibfnamefont {R.~T.}\ \bibnamefont
  {deSouza}}, \bibinfo {author} {\bibfnamefont {S.}~\bibnamefont {Hudan}},
  \bibinfo {author} {\bibfnamefont {V.~E.}\ \bibnamefont {Oberacker}},\ and\
  \bibinfo {author} {\bibfnamefont {A.~S.}\ \bibnamefont {Umar}},\ }\bibfield
  {title} {\bibinfo {title} {{C}onfronting measured near- and sub-barrier
  fusion cross sections for $^{20}\mathrm{O}+{}^{12}\mathrm{C}$ with a
  microscopic method},\ }\href {https://doi.org/10.1103/PhysRevC.88.014602}
  {\bibfield  {journal} {\bibinfo  {journal} {Phys. Rev. C}\ }\textbf {\bibinfo
  {volume} {88}},\ \bibinfo {pages} {014602} (\bibinfo {year}
  {2013})}\BibitemShut {NoStop}%
\bibitem [{\citenamefont {Umar}\ and\ \citenamefont
  {Oberacker}(2006)}]{Umar2006}%
  \BibitemOpen
  \bibfield  {author} {\bibinfo {author} {\bibfnamefont {A.~S.}\ \bibnamefont
  {Umar}}\ and\ \bibinfo {author} {\bibfnamefont {V.~E.}\ \bibnamefont
  {Oberacker}},\ }\bibfield  {title} {\bibinfo {title} {Heavy-ion interaction
  potential deduced from density-constrained time-dependent hartree-fock
  calculation},\ }\href {https://doi.org/10.1103/PhysRevC.74.021601} {\bibfield
   {journal} {\bibinfo  {journal} {Phys. Rev. C}\ }\textbf {\bibinfo {volume}
  {74}},\ \bibinfo {pages} {021601} (\bibinfo {year} {2006})}\BibitemShut
  {NoStop}%
\bibitem [{\citenamefont {Kortelainen}\ \emph {et~al.}(2012)\citenamefont
  {Kortelainen}, \citenamefont {McDonnell}, \citenamefont {Nazarewicz},
  \citenamefont {Reinhard}, \citenamefont {Sarich}, \citenamefont {Schunck},
  \citenamefont {Stoitsov},\ and\ \citenamefont {Wild}}]{kortelainen2012}%
  \BibitemOpen
  \bibfield  {author} {\bibinfo {author} {\bibfnamefont {M.}~\bibnamefont
  {Kortelainen}}, \bibinfo {author} {\bibfnamefont {J.}~\bibnamefont
  {McDonnell}}, \bibinfo {author} {\bibfnamefont {W.}~\bibnamefont
  {Nazarewicz}}, \bibinfo {author} {\bibfnamefont {P.-G.}\ \bibnamefont
  {Reinhard}}, \bibinfo {author} {\bibfnamefont {J.}~\bibnamefont {Sarich}},
  \bibinfo {author} {\bibfnamefont {N.}~\bibnamefont {Schunck}}, \bibinfo
  {author} {\bibfnamefont {M.~V.}\ \bibnamefont {Stoitsov}},\ and\ \bibinfo
  {author} {\bibfnamefont {S.~M.}\ \bibnamefont {Wild}},\ }\bibfield  {title}
  {\bibinfo {title} {Nuclear energy density optimization: Large deformations},\
  }\href {https://doi.org/10.1103/PhysRevC.85.024304} {\bibfield  {journal}
  {\bibinfo  {journal} {Phys. Rev. C}\ }\textbf {\bibinfo {volume} {85}},\
  \bibinfo {pages} {024304} (\bibinfo {year} {2012})}\BibitemShut {NoStop}%
\bibitem [{\citenamefont {McDonnell}\ \emph {et~al.}(2015)\citenamefont
  {McDonnell}, \citenamefont {Schunck}, \citenamefont {Higdon}, \citenamefont
  {Sarich}, \citenamefont {Wild},\ and\ \citenamefont
  {Nazarewicz}}]{mcdonnell2015}%
  \BibitemOpen
  \bibfield  {author} {\bibinfo {author} {\bibfnamefont {J.~D.}\ \bibnamefont
  {McDonnell}}, \bibinfo {author} {\bibfnamefont {N.}~\bibnamefont {Schunck}},
  \bibinfo {author} {\bibfnamefont {D.}~\bibnamefont {Higdon}}, \bibinfo
  {author} {\bibfnamefont {J.}~\bibnamefont {Sarich}}, \bibinfo {author}
  {\bibfnamefont {S.~M.}\ \bibnamefont {Wild}},\ and\ \bibinfo {author}
  {\bibfnamefont {W.}~\bibnamefont {Nazarewicz}},\ }\bibfield  {title}
  {\bibinfo {title} {Uncertainty quantification for nuclear density functional
  theory and information content of new measurements},\ }\href
  {https://doi.org/10.1103/PhysRevLett.114.122501} {\bibfield  {journal}
  {\bibinfo  {journal} {Phys. Rev. Lett.}\ }\textbf {\bibinfo {volume} {114}},\
  \bibinfo {pages} {122501} (\bibinfo {year} {2015})}\BibitemShut {NoStop}%
\bibitem [{\citenamefont {Godbey}\ \emph {et~al.}(2022)\citenamefont {Godbey},
  \citenamefont {Umar},\ and\ \citenamefont {Simenel}}]{godbey2022}%
  \BibitemOpen
  \bibfield  {author} {\bibinfo {author} {\bibfnamefont {K.}~\bibnamefont
  {Godbey}}, \bibinfo {author} {\bibfnamefont {A.~S.}\ \bibnamefont {Umar}},\
  and\ \bibinfo {author} {\bibfnamefont {C.}~\bibnamefont {Simenel}},\
  }\bibfield  {title} {\bibinfo {title} {Theoretical uncertainty quantification
  for heavy-ion fusion},\ }\href {https://doi.org/10.1103/PhysRevC.106.L051602}
  {\bibfield  {journal} {\bibinfo  {journal} {Phys. Rev. C}\ }\textbf {\bibinfo
  {volume} {106}},\ \bibinfo {pages} {L051602} (\bibinfo {year}
  {2022})}\BibitemShut {NoStop}%
\bibitem [{\citenamefont {Reinhard}\ \emph {et~al.}(2016)\citenamefont
  {Reinhard}, \citenamefont {Umar}, \citenamefont {Stevenson}, \citenamefont
  {Piekarewicz}, \citenamefont {Oberacker},\ and\ \citenamefont
  {Maruhn}}]{reinhard2016}%
  \BibitemOpen
  \bibfield  {author} {\bibinfo {author} {\bibfnamefont {P.-G.}\ \bibnamefont
  {Reinhard}}, \bibinfo {author} {\bibfnamefont {A.~S.}\ \bibnamefont {Umar}},
  \bibinfo {author} {\bibfnamefont {P.~D.}\ \bibnamefont {Stevenson}}, \bibinfo
  {author} {\bibfnamefont {J.}~\bibnamefont {Piekarewicz}}, \bibinfo {author}
  {\bibfnamefont {V.~E.}\ \bibnamefont {Oberacker}},\ and\ \bibinfo {author}
  {\bibfnamefont {J.~A.}\ \bibnamefont {Maruhn}},\ }\bibfield  {title}
  {\bibinfo {title} {Sensitivity of the fusion cross section to the density
  dependence of the symmetry energy},\ }\href
  {https://doi.org/10.1103/PhysRevC.93.044618} {\bibfield  {journal} {\bibinfo
  {journal} {Phys. Rev. C}\ }\textbf {\bibinfo {volume} {93}},\ \bibinfo
  {pages} {044618} (\bibinfo {year} {2016})}\BibitemShut {NoStop}%
\bibitem [{\citenamefont {Schuetrumpf}\ and\ \citenamefont
  {Nazarewicz}(2017)}]{Schuetrumpf2017}%
  \BibitemOpen
  \bibfield  {author} {\bibinfo {author} {\bibfnamefont {B.}~\bibnamefont
  {Schuetrumpf}}\ and\ \bibinfo {author} {\bibfnamefont {W.}~\bibnamefont
  {Nazarewicz}},\ }\bibfield  {title} {\bibinfo {title} {Cluster formation in
  precompound nuclei in the time-dependent framework},\ }\href
  {https://doi.org/10.1103/PhysRevC.96.064608} {\bibfield  {journal} {\bibinfo
  {journal} {Phys. Rev. C}\ }\textbf {\bibinfo {volume} {96}},\ \bibinfo
  {pages} {064608} (\bibinfo {year} {2017})}\BibitemShut {NoStop}%
\bibitem [{sup()}]{sup}%
  \BibitemOpen
  \href@noop {} {}\bibinfo {note} {See Supplemental Material at [URL inserted
  by publisher] for videos of the time evolution of the neutron localization
  function for the collision of $^{16,17}$O + $^{12}$C and a figure of the
  isovector part of the DC-TDHF potential for the $^{16,17,18}$O+$^{12}$C
  reaction.}\BibitemShut {Stop}%
\bibitem [{\citenamefont {Asher}\ \emph {et~al.}(2021)\citenamefont {Asher},
  \citenamefont {Almaraz-Calderon}, \citenamefont {Kemper}, \citenamefont
  {Baby}, \citenamefont {Lopez-Saavedra}, \citenamefont {Morelock},
  \citenamefont {Perello}, \citenamefont {Tripathi},\ and\ \citenamefont
  {Keeley}}]{Asher21a}%
  \BibitemOpen
  \bibfield  {author} {\bibinfo {author} {\bibfnamefont {B.}~\bibnamefont
  {Asher}}, \bibinfo {author} {\bibfnamefont {S.}~\bibnamefont
  {Almaraz-Calderon}}, \bibinfo {author} {\bibfnamefont {K.}~\bibnamefont
  {Kemper}}, \bibinfo {author} {\bibfnamefont {L.}~\bibnamefont {Baby}},
  \bibinfo {author} {\bibfnamefont {E.}~\bibnamefont {Lopez-Saavedra}},
  \bibinfo {author} {\bibfnamefont {A.}~\bibnamefont {Morelock}}, \bibinfo
  {author} {\bibfnamefont {J.}~\bibnamefont {Perello}}, \bibinfo {author}
  {\bibfnamefont {V.}~\bibnamefont {Tripathi}},\ and\ \bibinfo {author}
  {\bibfnamefont {N.}~\bibnamefont {Keeley}},\ }\bibfield  {title} {\bibinfo
  {title} {Resolution of a long-standing discrepancy in the
  $^{17}${O}+$^{12}${C} fusion excitation function},\ }\href
  {https://doi.org/https://doi.org/10.1140/epja/s10050-021-00584-8} {\bibfield
  {journal} {\bibinfo  {journal} {The European Physical Journal A}\ }\textbf
  {\bibinfo {volume} {57}},\ \bibinfo {pages} {272} (\bibinfo {year}
  {2021})}\BibitemShut {NoStop}%
\bibitem [{\citenamefont {Tighe}\ \emph {et~al.}(1993)\citenamefont {Tighe},
  \citenamefont {Kolata}, \citenamefont {Belbot},\ and\ \citenamefont
  {E.F.}}]{Tighe93}%
  \BibitemOpen
  \bibfield  {author} {\bibinfo {author} {\bibfnamefont {R.}~\bibnamefont
  {Tighe}}, \bibinfo {author} {\bibfnamefont {J.}~\bibnamefont {Kolata}},
  \bibinfo {author} {\bibfnamefont {M.}~\bibnamefont {Belbot}},\ and\ \bibinfo
  {author} {\bibfnamefont {A.}~\bibnamefont {E.F.}},\ }\bibfield  {title}
  {\bibinfo {title} {{P}ossible signatures of nuclear-molecular formation in
  {O}+{C} systems},\ }\href {https://doi.org/10.1103/PhysRevC.47.2699}
  {\bibfield  {journal} {\bibinfo  {journal} {Phys. Rev. C}\ }\textbf {\bibinfo
  {volume} {47}},\ \bibinfo {pages} {2699} (\bibinfo {year}
  {1993})}\BibitemShut {NoStop}%
\bibitem [{\citenamefont {A.~Hertz}\ \emph {et~al.}(1978)\citenamefont
  {A.~Hertz}, \citenamefont {Essel}, \citenamefont {Körner}, \citenamefont
  {Rehm},\ and\ \citenamefont {Sperr}}]{Hertz78}%
  \BibitemOpen
  \bibfield  {author} {\bibinfo {author} {\bibfnamefont {A.}~\bibnamefont
  {A.~Hertz}}, \bibinfo {author} {\bibfnamefont {H.}~\bibnamefont {Essel}},
  \bibinfo {author} {\bibfnamefont {H.~J.}\ \bibnamefont {Körner}}, \bibinfo
  {author} {\bibfnamefont {K.~E.}\ \bibnamefont {Rehm}},\ and\ \bibinfo
  {author} {\bibfnamefont {P.}~\bibnamefont {Sperr}},\ }\bibfield  {title}
  {\bibinfo {title} {Maximum fusion cross section for the system $^{12}${C} +
  $^{17}${O}},\ }\href {https://doi.org/10.1103/PhysRevC.18.2780} {\bibfield
  {journal} {\bibinfo  {journal} {Phys. Rev. C}\ }\textbf {\bibinfo {volume}
  {18}},\ \bibinfo {pages} {2780(R)} (\bibinfo {year} {1978})}\BibitemShut
  {NoStop}%
\bibitem [{\citenamefont {Hofmann}\ \emph {et~al.}(1997)\citenamefont
  {Hofmann}, \citenamefont {He{\ss}berger}, \citenamefont {Ninov},
  \citenamefont {Armbruster}, \citenamefont {M{\"u}nzenberg}, \citenamefont
  {Stodel}, \citenamefont {Popeko}, \citenamefont {Yeremin}, \citenamefont
  {Saro},\ and\ \citenamefont {Leino}}]{Hofmann1997}%
  \BibitemOpen
  \bibfield  {author} {\bibinfo {author} {\bibfnamefont {S.}~\bibnamefont
  {Hofmann}}, \bibinfo {author} {\bibfnamefont {F.~P.}\ \bibnamefont
  {He{\ss}berger}}, \bibinfo {author} {\bibfnamefont {V.}~\bibnamefont
  {Ninov}}, \bibinfo {author} {\bibfnamefont {P.}~\bibnamefont {Armbruster}},
  \bibinfo {author} {\bibfnamefont {G.}~\bibnamefont {M{\"u}nzenberg}},
  \bibinfo {author} {\bibfnamefont {C.}~\bibnamefont {Stodel}}, \bibinfo
  {author} {\bibfnamefont {A.~G.}\ \bibnamefont {Popeko}}, \bibinfo {author}
  {\bibfnamefont {A.~V.}\ \bibnamefont {Yeremin}}, \bibinfo {author}
  {\bibfnamefont {S.}~\bibnamefont {Saro}},\ and\ \bibinfo {author}
  {\bibfnamefont {M.}~\bibnamefont {Leino}},\ }\bibfield  {title} {\bibinfo
  {title} {Excitation function for the production of $^{265}$108 and
  $^{266}$109},\ }\href {https://doi.org/10.1007/s002180050343} {\bibfield
  {journal} {\bibinfo  {journal} {Z. Phys. A}\ }\textbf {\bibinfo {volume}
  {358}},\ \bibinfo {pages} {377} (\bibinfo {year} {1997})}\BibitemShut
  {NoStop}%
\bibitem [{\citenamefont {Heusch}\ \emph {et~al.}(1982)\citenamefont {Heusch},
  \citenamefont {Beck}, \citenamefont {Coffin}, \citenamefont {Engelstein},
  \citenamefont {Freeman}, \citenamefont {Guillaume}, \citenamefont {Haas},\
  and\ \citenamefont {Wagner}}]{Heusch82}%
  \BibitemOpen
  \bibfield  {author} {\bibinfo {author} {\bibfnamefont {B.}~\bibnamefont
  {Heusch}}, \bibinfo {author} {\bibfnamefont {C.}~\bibnamefont {Beck}},
  \bibinfo {author} {\bibfnamefont {J.~P.}\ \bibnamefont {Coffin}}, \bibinfo
  {author} {\bibfnamefont {P.}~\bibnamefont {Engelstein}}, \bibinfo {author}
  {\bibfnamefont {R.~M.}\ \bibnamefont {Freeman}}, \bibinfo {author}
  {\bibfnamefont {G.}~\bibnamefont {Guillaume}}, \bibinfo {author}
  {\bibfnamefont {F.}~\bibnamefont {Haas}},\ and\ \bibinfo {author}
  {\bibfnamefont {P.}~\bibnamefont {Wagner}},\ }\bibfield  {title} {\bibinfo
  {title} {Entrance channel effect for complete fusion of {O} + {C} isotopes},\
  }\href {https://doi.org/10.1103/PhysRevC.26.542} {\bibfield  {journal}
  {\bibinfo  {journal} {Phys. Rev. C}\ }\textbf {\bibinfo {volume} {26}},\
  \bibinfo {pages} {542} (\bibinfo {year} {1982})}\BibitemShut {NoStop}%
\bibitem [{\citenamefont {Magierski}\ \emph {et~al.}(2017)\citenamefont
  {Magierski}, \citenamefont {Sekizawa},\ and\ \citenamefont
  {Wlaz\l{}owski}}]{Magierski2017}%
  \BibitemOpen
  \bibfield  {author} {\bibinfo {author} {\bibfnamefont {P.}~\bibnamefont
  {Magierski}}, \bibinfo {author} {\bibfnamefont {K.}~\bibnamefont
  {Sekizawa}},\ and\ \bibinfo {author} {\bibfnamefont {G.}~\bibnamefont
  {Wlaz\l{}owski}},\ }\bibfield  {title} {\bibinfo {title} {Novel role of
  superfluidity in low-energy nuclear reactions},\ }\href
  {https://doi.org/10.1103/PhysRevLett.119.042501} {\bibfield  {journal}
  {\bibinfo  {journal} {Phys. Rev. Lett.}\ }\textbf {\bibinfo {volume} {119}},\
  \bibinfo {pages} {042501} (\bibinfo {year} {2017})}\BibitemShut {NoStop}%
\end{thebibliography}
%

\end{document}


\title{Supplemental Material for 
``In search of beyond mean-field signatures in heavy-ion fusion reactions''}

\author{R.~T. deSouza\,\orcidlink{0000-0001-5835-677X}}\email{desouza@indiana.edu}{\altaffiliation{
Department of Chemistry and Center for Exploration of Energy and Matter, Indiana University, Bloomington, Indiana 47408, USA}

\author{K. Godbey\,\orcidlink{0000-0003-0622-3646}} 
\email{godbey@frib.msu.edu}\altaffiliation{
FRIB Laboratory, Michigan State University, East Lansing, Michigan 48824, USA} 

\author{S. Hudan\,\orcidlink{0000-0002-9722-2245}}
\email{shudan@indiana.edu}{\altaffiliation{
Department of Chemistry and Center for Exploration of Energy and Matter, Indiana University\, Bloomington, Indiana 47408, USA}

\author{W. Nazarewicz\,\orcidlink{0000-0002-8084-7425}}\email{witek@frib.msu.edu}\altaffiliation{Department of Physics and Astronomy and FRIB Laboratory, Michigan State University, East Lansing, Michigan 48824, USA}

\begin{abstract}
This supplemental material contains 2 illustrative videos of TDHF simulations and one supplemental figure. 
\end{abstract}

\maketitle
\setcounter{figure}{0}
\renewcommand{\thefigure}{S\arabic{figure}}
\renewcommand{\thesection}{S.\Roman{section}}

\section{Animations of TDHF simulations}

\noindent
{\bf \href{https://drive.google.com/file/d/1I8IZusnA9YTiyOvynCzmFPu_MIqAe0no/view?usp=drive_link}{Video 1}}: Time evolution of the neutron localization function for the collision of $^{16,17}$O + $^{12}$C at $E_{\rm c.m.}$ = 14 MeV and the incoming angular momentum $\ell = 0 \hbar$ as predicted by TDHF calculations.  \\

\noindent
{\bf \href{https://drive.google.com/file/d/1Otamt3B4hB2abbQNQqL_e7LJytvnsmMa/view?usp=drive_link}{Video 2}}: Time evolution of the neutron localization function for the collision of $^{16,17}$O + $^{12}$C at $E_{\rm c.m.}$ = 14 MeV and the incoming angular momentum $\ell = 11 \hbar$ as predicted by TDHF calculations.  \\


\section{DC-TDHF isovector potential}

\begin{figure}[!htb]
\includegraphics[width=1.\linewidth]{Isovector.pdf}
\caption{The isovector part of the DC-TDHF potential for the $^{16,17,18}$O$+^{12}$C reactions discussed in the main text. For $^{17}$O, 200 random initial orientations of the deformed projectile were considered and are plotted as the light grey lines.}
\label{fig:IV}
\end{figure}